\documentclass[preprint]{aastex61}

\begin{document}

\title{3XMM J181923.7$-$170616: an X-ray binary with a 408~s
pulsar}

\newcommand{\totder}[2]{\frac{d{#1}}{d{#2}}}

\newcommand{\gsim}{\mbox{\hspace{.2em}\raisebox{.5ex}{$>$}\hspace{-.8em}\raisebox{-.5ex}{$\sim$}\hspace{.2em}}}
\newcommand{\lsim}{\mbox{\hspace{.2em}\raisebox{.5ex}{$<$}\hspace{-.8em}\raisebox{-.5ex}{$\sim$}\hspace{.2em}}}
\newcommand{\ssst}{\scriptscriptstyle}
\newcommand{\E}[1]{\times 10^{#1}}
\newcommand{\etal}{et al.}
\newcommand{\lt}{\left}       \newcommand{\rt}{\right}
\newcommand{\RA}[3]{{#1}^{{\rm h}}{#2}^{{\rm m}}{#3}^{\rm s}}
\newcommand{\decl}[3]{{#1}^{\circ}{#2}'{#3}''}
\newcommand{\RAdot}[4]{{#1}^{{\rm h}}{#2}^{{\rm m}}{#3}\fs{#4}}
\newcommand{\decldot}[4]{{#1}^{\circ}{#2}'{#3}\farcs{#4}}

\newcommand{\s}{\,{\rm s}}      \newcommand{\ps}{\,{\rm s}^{-1}}
\newcommand{\yr}{\,{\rm yr}}    \newcommand{\Msun}{M_{\odot}}
\newcommand{\cm}{\,{\rm cm}}    \newcommand{\km}{\,{\rm km}}
\newcommand{\kms}{$\km\ps$}
\newcommand{\parsec}{\,{\rm pc}}\newcommand{\kpc}{\,{\rm kpc}} \newcommand{\pc}{\,{\rm pc}}
\newcommand{\erg}{\,{\rm erg}}        \newcommand{\K}{\,{\rm K}}
\newcommand{\eV}{\,{\rm eV}}    \newcommand{\keV}{\,{\rm keV}}
\newcommand{\Hz}{\,{\rm Hz}}    
\newcommand{\as}{$^{\prime\prime}\ $} \newcommand{\am}{$^{\prime}\ $}
\newcommand{\um}{\,\mu\rm m}
\newcommand{\degree}{\ensuremath{^\circ}}

\newcommand{\nel}{n_{e}}        \newcommand{\NH}{N_{\ssst\rm H}}
\newcommand{\no}{n_{\ssst 0}}
\newcommand{\Ts}{T_{s}}
\newcommand{\kTc}{kT_{\rm c}} \newcommand{\kTh}{kT_{\rm h}}
\newcommand{\Tc}{T_{\rm c}} \newcommand{\Th}{T_{\rm h}}
\newcommand{\tauc}{\tau_{\rm c}} \newcommand{\tauh}{\tau_{\rm h}}
\newcommand{\rs}{r_{s}}         \newcommand{\vs}{v_{s}}
\newcommand{\nH}{n_{\ssst\rm H}}        \newcommand{\mH}{m_{\ssst\rm H}}
\newcommand{\nHH}{n({\rm H}_{2})} \newcommand{\NHH}{N({\rm H}_{2})}
\newcommand{\MHH}{M({\rm H}_{2})}
\newcommand{\VLSR}{V_{\ssst\rm LSR}}
\newcommand{\Tmb}{T_{\rm mb}}
\newcommand{\xray}{X-ray}       \newcommand{\Einstein}{{\sl Einstein}}
\newcommand{\ROSAT}{{\sl ROSAT}} \newcommand{\Spitzer}{{\sl Spitzer}}
\newcommand{\FCRAO}{{\sl FCRAO}}
\newcommand{\XMMN}{{\sl XMM-Newton}}
\newcommand{\Chandra}{{\sl Chandra}}
\newcommand{\ASCA}{{\sl ASCA}}
\newcommand{\Suzaku}{{\sl Suzaku}}
\newcommand{\Swift}{{\sl Swift}}
\newcommand{\Fermi}{{\sl Fermi}}
\newcommand{\du}{d_{7.1}} \newcommand{\Du}{D_{20}}
\newcommand{\dg}{d_{10}} 
\newcommand{\Eu}{E_{51}}    \newcommand{\ru}{R_{12}}
\newcommand{\Rb}{R_{\rm b}}
\newcommand{\mE}{\langle E_{\ssst B-V}\rangle}
\newcommand{\rray}{$\gamma$-ray}
\newcommand{\Ha}{H$\alpha$}
\newcommand{\Hea}{He$\alpha$}
\newcommand{\Lya}{Ly$\alpha$}
\newcommand{\Ka}{K$\alpha$}
\newcommand{\SII}{S~{\footnotesize II}}

\newcommand{\twCO}{$^{12}$CO}   \newcommand{\thCO}{$^{13}$CO}
\newcommand{\HCOp}{HCO$+$}
\newcommand{\Jotz}{$J$=1--0}    \newcommand{\Jtto}{$J$=2--1}
\newcommand{\Jttt}{$J$=3--2}
\newcommand{\vw}{v_{\rm w}}

\newcommand{\vmekal}{$vmekal$}
\newcommand{\cie}{$cie$}
\newcommand{\nei}{$nei$}
\newcommand{\pshock}{$pshock$}
\newcommand{\vnei}{$vnei$}
\newcommand{\apec}{$apec$}
\newcommand{\vpshock}{$vpshock$}
\newcommand{\vsedov}{$vsedov$}
\newcommand{\powerlaw}{$power$-$law$}

\newcommand{\HI}{\ion{H}{1}}


\author{Hao Qiu}
\affil{School of Astronomy and Space Science, Nanjing University,
163 Xianlin Avenue, Nanjing, 210023, China} 

\author{Ping Zhou}
\affil{School of Astronomy and Space Science, Nanjing University,
163 Xianlin Avenue, Nanjing, 210023, China}
\affil{Anton Pannekoek Institute for Astronomy, University of Amsterdam, 
Science Park 904, 1098 XH Amsterdam, The Netherlands}
\altaffiliation{p.zhou@uva.nl}
\affil{Key Laboratory of Modern Astronomy and Astrophysics,
Nanjing University, Ministry of Education, China}

\author{Wenfei Yu}
\affiliation{Key Laboratory for Research in Galaxies and Cosmology, Shanghai Astronomical Observatory, 
Chinese Academy of Sciences, 80 Nandan Road, Shanghai 200030, China}

\author{Xiangdong Li}
\affil{School of Astronomy and Space Science, Nanjing University, 
163 Xianlin Avenue, Nanjing, 210023, China}
\affil{Key Laboratory of Modern Astronomy and Astrophysics,
Nanjing University, Ministry of Education, China}

\author{Xiaojie Xu}
\affil{School of Astronomy and Space Science, Nanjing University,
163 Xianlin Avenue, Nanjing, 210023, China}
\affil{Key Laboratory of Modern Astronomy and Astrophysics,
Nanjing University, Ministry of Education, China}



\begin{abstract}

We carry out a dedicated study of 3XMM J181923.7$-$170616\ with an approximate pulsation 
period of 400~s { using the \XMMN\ and \Swift\ observations spanning 
across nine years.
We have refined the period of the source to 407.904(7) s (at epoch MJD
57142) and {determined a period derivative limit of $\dot{P}{\leq} 5.9\pm 5.4\E{-9} \s\ps$ ($1\sigma$).}
}
The source radiates hard, persistent X-ray emission during the 
observation epochs, which is best described by an absorbed 
\powerlaw\ model ($\Gamma \sim 0.2$--0.8) plus faint Fe lines at 
6.4 keV and 6.7 keV. 
{  The X-ray flux revealed a variation within a factor of 2,
along with a spectral hardening as the
flux increased.}
The pulse shape is sinusoid-like and the spectral properties
of different phases do not present significant variation.
The absorption $\NH$ ($\sim 1.3\E{22}~\cm^{-2}$) is similar to
the total Galactic hydrogen column density along the direction, 
indicating that it is a distant source.
A search for the counterpart in optical and near-infrared surveys reveals 
a low mass K-type giant, while the existence of a Galactic 
OB supergiant is excluded. 
A symbiotic X-ray binary is the favored nature of 3XMM J181923.7$-$170616\ and can essentially 
explain the low luminosity of $2.78\E{34} \dg^{2} \erg\ps$, 
slow pulsation, hard X-ray spectrum, and possible K3~III companion.
An alternative explanation of the source is a persistent Be/X-ray binary
with a companion star no earlier than B3-type.

\end{abstract}




\section{Introduction} \label{sec:intro}

X-ray Binaries (XRBs) are binary systems in which a compact object 
(such as a neutron star or a black hole) is accreting from a companion star, 
contributing a significant amount of X-ray radiation in one galaxy.
XRBs are divided into two categories according to the mass of the companion
star. 
A binary with a massive companion ($ > 8 \Msun$) is called a 
high mass X-ray binary (HMXB),while those with smaller 
companions ($<1 \Msun$) are classified as low mass X-ray binaries (LMXBs).

LMXBs are usually transient sources, in their quiescent states the X-ray luminosities
can be down to 10$^{32} \erg \s^{-1}$ in the soft X-ray band,
and they can turn to be very luminous during the outbursts (up to 10$^{38}$ erg 
s$^{-1}$, see Reig 2011 for a recent review). 
HMXB systems are mainly subdivided into Be/X-ray Binaries (BeXBs) and 
supergiant X-ray binaries (SGXBs). 
BeXBs contain a neutron star (NS) orbiting a Be star (Negueruela 1998).
Most of them display two types of outbursts (Type I and II) during which the 
luminosity is increased significantly ($\Delta L_X\sim $100-1000).
With the improved sensibilities of X/$\gamma$-ray telescopes,
the above picture of XRBs has been enriched with emerging subclasses.
Symbiotic X-ray binaries (SyXBs) are a new subclass of LMXBs
containing an X-ray luminous NS and a late type giant companion.
Our current knowledge of SyXBs comes from about 10 SyXBs or potential candidates,
which manifest long orbital and spin periods ($> 100$ s) and a relatively low X-ray 
luminosity ($10^{32}$--$10^{36} \erg\ps$; L\"{u} \etal\ 2012).
The long spin period and weak X-ray emission are also typical for
the persistent BeXBs, in which the NS is suggested to be
far from the Be star and accreting material from 
the low-density regions of the Be star's wind (Reig \& Roche 1999).

During the past two decades, the X-ray space telescopes \XMMN\ and \Chandra\ have
detected numerous serendipitous X-ray sources thanks to their high sensibilities.
Their catalogues record X-ray objects with a variety of populations
and provide rich resources for the exploration of new XRBs.

3XMM J181923.7$-$170616 is an X-ray source recorded in the third XMM serendipitous source 
catalogue Data Release 5 (3XMM-DR5; Rosen et al.\ 2016).
It was identified as a slow X-ray pulsar candidate with a period of 400~s in a 
study of the 3XMM-DR4 using the machine learning method (Farrell et al.\ 2015).
We performed periodicity search of a selected sample of sources in 3XMM-DR5 and
also noticed the clear periodicity of the source at $\sim 408~\s$.

To uncover the nature of 3XMM J181923.7$-$170616,
we carried out a dedicated analysis of 3XMM J181923.7$-$170616\ using 12 new \Swift\
observations in addition to the \XMMN\ observations.
In Section~2 we describe the X-ray observations used in this study.
Our timing and spectral analysis with the \XMMN\ and \Swift-XRT observations 
are presented in Section 3. 
The search for the counterpart and our discussion related to the nature 
are included in Section 4. Section 5 summarizes the results.

\section{Observations} \label{sec:observ}
\subsection{\XMMN}\label{subsec:obs-xmm}

3XMM J181923.7$-$170616\ was observed in three \XMMN\ observations with the European Photon Imaging 
Camera (EPIC), which contains two MOS cameras (Turner \etal\ 2001) and 
a pn camera (Str\"{u}der \etal\ 2001) to detect X-ray photons in 
the 0.2--10 keV energy range. 
All of these three observations were originally pointed to an HXMB SAX J1818.6-1703 ($\RAdot{18}{18}{37}{90}$, $\decldot{-17}{02}{47}{96}$, J2000) and 
detected the serendipitous bright source 3XMM J181923.7$-$170616\ $12'$
away to the southeast.
Two observations of the source conducted in 2006 (PI D.M. Smith) and 
2010 (PI E. Bozzo) were recorded in 3XMM-DR5. 
A third observation (PI S. Drave) was carried out in 2013, it 
was not included in the 3XMM catalogue.
We only use the MOS data of the third observation because 3XMM J181923.7$-$170616\ was 
beyond the field of view of the pn camera observation. 
{ The Reflection Grating Spectrometer (RGS) data of each observation 
has a cross dispersion of $5'$, in which the source was out of the field.
Therefore we do not perform an analysis of the RGS data.}
The time resolutions of the MOS and pn data are 2.6~s and 73.4~ms, 
respectively.
We removed the time intervals with strong heavy proton flaring by 
checking the CCD corner light curves.
Since the third observation has an unsteady light curve in most
of the observation time, we only used the time intervals with
flat, low counts rate and did not refer to this observation for 
precise analysis.
The total screened exposures for the MOS1/2 and pn data are 
54 ks and 26.7 ks, respectively.
XMM Science Analysis System software (SAS, ver 15.0) was used to 
reproduce the \XMMN\ data.


\subsection{\Swift}\label{subsec:obs-swift}

We found additional data of 3XMM J181923.7$-$170616\ observed with the \Swift\ space telescope
in the HEASARC\footnote{http://heasarc.gsfc.nasa.gov/} data archive.
There are 12 observations toward the X-ray point source SWIFT J1819.2-1706 that 
spatially matches 3XMM J181923.7$-$170616. We believe that they are the same source because of their
spatial coincidence and identical timing and spectral properties (see 
Section~\ref{S:results}).

The \Swift\ observations were conducted between April 30 and May 31, 2015. 
We retrieved the Ultraviolet/Optical Telescope (UVOT) imaging data for the 170-650 nm 
and X-ray telescope (XRT) data in the 0.2-10 keV energy range. 
We chose the PC mode of XRT data and processed them using the HEASOFT Calibration Database.
UVOT data are not used in this paper, because we did not find a counterpart 
of the source in the \Swift\ UV/Optical image.
The XRT CCD time resolution is 2.5 s.
The minimum, maximum, and total exposures amongst XRT observations are
approximately 1.5 ks, 10.3 ks, and 88.3 ks, respectively.

XSPEC(ver 12.9.0) and XRONOS packages in HEASOFT(ver 6.17), and 
TEMPO2\footnote{http://www.atnf.csiro.au/research/pulsar/tempo2/} (Hobbs \etal\ 2006; Edwards \etal\ 2006) 
were used for timing and spectral analysis of the \XMMN\ and \Swift\ data.
We also used Python software packages (Astropy, Scipy/Numpy and Matplotlib)
 for data analysis and visualization. 
Table~\ref{T:obs} lists the detailed information (the observation ID, date, exposure, 
and average count rate) of the X-ray observations.

\section{Results} \label{S:results}

\subsection{Timing Analysis} \label{sec:timing}
The photons of 3XMM J181923.7$-$170616\ are extracted from a circular region with a radius 
of $30"$ centered at ($\RAdot{18}{19}{23}{77}$, 
$\decldot{-17}{06}{15}{25}$, J2000).
We corrected the photon's arrival time to the barycenter of the solar 
system before the periodicity search.
The power spectral density of the time series in 0.2--10 keV band
was calculated with the \textit{powspec} command in the XRONOS package.
The upper limits of the calculated frequencies are $\sim 0.2$ $\Hz$ for
the MOS and XRT data, and 7~$\Hz$ for the pn data.
A power density peak is shown at 0.0025 Hz in the \XMMN\ and \Swift\
power spectra, corresponding to a periodicity of $\sim 400\s$.
We subsequently apply the epoch-folding method (\textit{efsearch}; Leahy 1987),
which refines the periodicity to around 408~s.
{ The uncertainty of the period $\sigma_{p}$ is
estimated with $\sigma_{P} = 0.71(\chi_{r}^{2}-1)^{-0.63} \Delta P$,
where $\Delta P=P^2/(2T_s)$ is the Fourier resolution for an observation
duration of $T_s$ and $\chi_{r}^{2}$ is the peak theoretical reduced 
Chi-square in the $\chi^2$ vs. period plot (see Leahy 1987).}
Table~\ref{T:obs} summarizes the 9 groups (3 for MOS1, 3 for MOS2, 2 for pn, 
and 1 for XRT of \Swift) of periods.
In this process we combine the 12 \Swift\ observations taken within two months. 

{ We construct a phase-connection analysis by fitting the time-of-arrivals
(TOAs) of the observations to a standard timing model:
$\phi(t) = \phi(t_0)+ (t-t_0)/P-(t-t_0)^2\dot{P}/(2P^2)$.
The TOA of each observation is determined by folding the 
pulse profile with a period of 407.9091 s. 
This period with a small uncertainty (0.0025 s) is taken from the 
epoch-folding result of the \Swift\ data (see Table~\ref{T:obs}).
A sinusoid can reproduce each folded pulse profile and is thus used 
to determine the TOA of each observation. 
The \Swift\ data with the ID of 00033498008 are not used for the phase connection, 
since the short exposure (1.5 ks; less than 4 periods) is insufficient to provide a reliable TOA.

\begin{deluxetable}{lcclcc}
\tabletypesize{\footnotesize}
\tablecaption{Summary of the \XMMN\ and \Swift\ observations and
  the timing properties of 3XMM J181923.7$-$170616\ }
    \tablewidth{0pt}
    \tablehead{
      \colhead{Obs. ID}  & Obs. Date & Exposure$^a$ & Instrument & Count Rate$^{b}$
        & Period$^{c}$ \\
          & & (ks) &  & (counts $\s^{-1}$) & (s) 
          }
          \startdata
          0402470101 &  2006 Oct 07     & 13.1/18.2     & XMM/MOS1      & 0.036(2)      & 408.0(9)      \\
          & &                   13.3/18.2               & XMM/MOS2      & 0.037(2)      & 409(1)        \\
          & &                   11.7/18.2               & XMM/pn        & 0.058(4)      & 408.9(6)      \\
          0604820101    & 2010 Mar 21   & 28.6/45.6     & XMM/MOS1      & 0.045(2)      & 407.7(2)      \\
          & &                   24.3/45.6               & XMM/MOS2      & 0.042(2)      & 408.2(2)      \\
          & &                   15.0/45.6               & XMM/pn                &0.108(4)       & 407.7(2)         \\
          0693900101$^d$        & 2013 Mar 21   & 12.5/30.7     & XMM/MOS1      & 0.054(3)      & 406.7(3)  \\
          & &                   12.5/30.7               & XMM/MOS2      & 0.054(2)      & 409.1(4)      \\     
          00033498001   & 2015 Apr 30   & 9.6   & Swift/XRT     & 0.022(1)      & 407.909(3) \\
          00033498002   & 2015 May 01   & 8.6   & Swift/XRT     & 0.026(1)      &-  \\
          00033498003   & 2015 May 02   & 5.0   & Swift/XRT     & 0.027(2)& -\\
          00033498004   & 2015 May 03   & 9.8   & Swift/XRT     &0.026(1)& -\\
          00033498005   & 2015 May 04   & 10.3  & Swift/XRT     &0.022(1)&- \\
          00033498006   & 2015 May 05   & 9.0   & Swift/XRT     & 0.019(1)& -\\
          00033498007   & 2015 May 11   & 7.3   & Swift/XRT     & 0.024(1)& -\\
          00033498008   & 2015 May 14   & 1.5   & Swift/XRT     & 0.024(4)& -\\
          00033498009   & 2015 May 16   & 5.5   & Swift/XRT     &0.018(2)& -\\
          00033498010   & 2015 May 21   & 9.0   & Swift/XRT     & 0.026(1)& -\\
          00033498011   & 2015 May 27   & 5.2   & Swift/XRT     & 0.019(3)& -\\
          00033498012   & 2015 May 31   & 7.5   & Swift/XRT     &0.019(2)& -\\
\enddata
\tablenotetext{a}{\phantom{0} For \XMMN\ observations we show the flare-screened exposure/the total exposure}
\tablenotetext{b}{\phantom{0} The count rate in 0.3--10 keV is calculated 
in the flare-screened time. 
The 1-$\sigma$ uncertainty of the last digit is given in the parentheses.}
\tablenotetext{c}{\phantom{0} The uncertainty (Leahy 1987) of the last 
one digit is given in parentheses.}
\tablenotetext{d}{\phantom{0} The observation suffered flarings in most of the observation time, which may cause problematic
timing results.}
\label{T:obs}
\end{deluxetable}
\begin{figure}[tbh!]
\centering
\includegraphics[angle=0, width=0.9\textwidth]{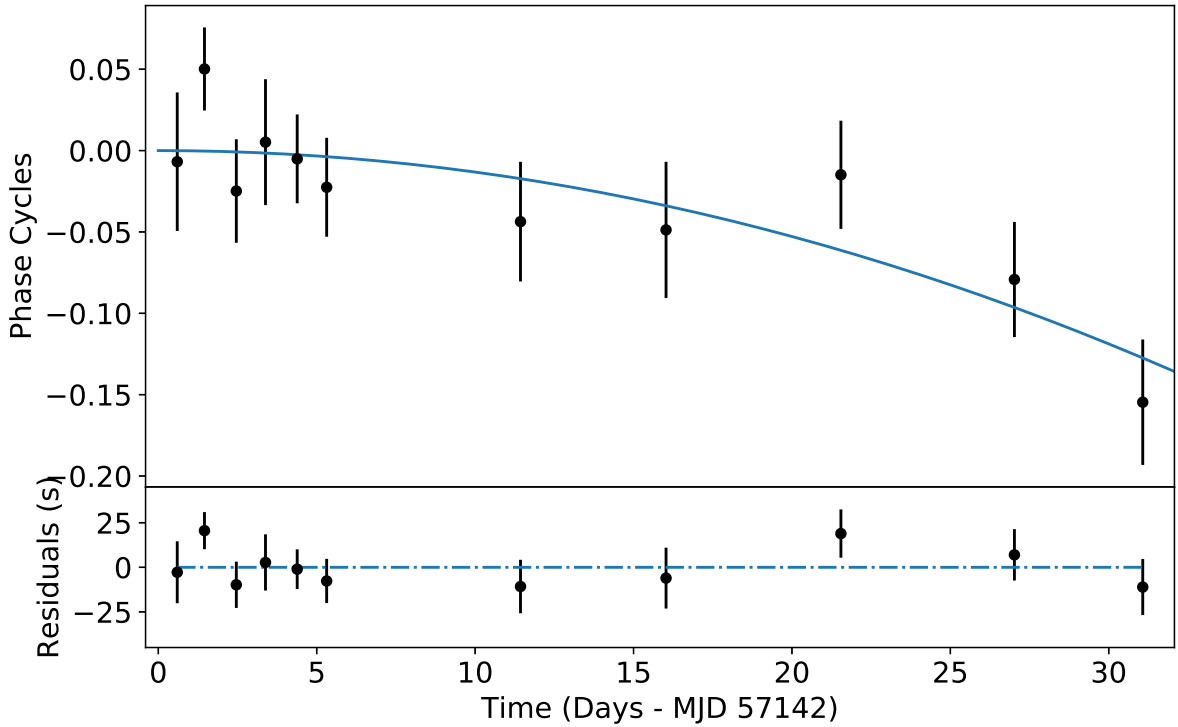}
\caption{Time/Phase residuals of the 11 XRT TOAs 
after subtraction of the best-fit quadratic ephemeris model 
($P=407.904$ s at MJD 57142, $\dot{P}=5.9\times10^{-9} $s s$^{-1}$).
The top panel shows the timing residuals subtracting only the 
contribution of the linear component of the best-fit ephemeris, 
with the solid line showing the quadratic term.
The lower panel gives the time residuals respect to the 
quadratic ephemeris model.}
\end{figure}
\label{f:phase-connection}

We started by fitting the first three TOAs to a linear ephemeris 
and added TOAs one at a time using a quadratic ephemeris.
We repeated the fit procedure iteratively using the new solution until the
last TOA is added.
Each newly added TOA matches to $\lesssim 0.1$ cycles of the predicted phase derived from 
the previous solution.
The best-fit quadratic ephemeris is given by the final 
phase-fitting procedure:
$P=407.904(7)~\s$ on MJD 57142 and $\dot{P}=5.9\pm 5.4\E{-9} \s\ps$ (1$\sigma$ uncertainty; 
rms=10.8 s, $\chi^2/d.o.f=8.2/8=1.03$; see Figure~\ref{f:phase-connection}. 
). 
During the Swift observation epoch, the quadratic term 
contributes -0.06 (-0.01 -- -0.24) cycles.
We did not fit retroactively back to the \XMMN\ TOAs, considering that a constant $\dot{f}$ 
would contribute  -37 -- -882 cycles between the last  \XMMN\ TOA and the first \Swift\ TOA.
The quadratic solution is marginally better than a linear solution ($P=407.912(2)$;  rms=11.5~s, $\chi^2/d.o.f=9.6/9=1.06$).

\begin{figure}[tbh!]
\centering
\includegraphics[angle=0, width=0.5\textwidth]{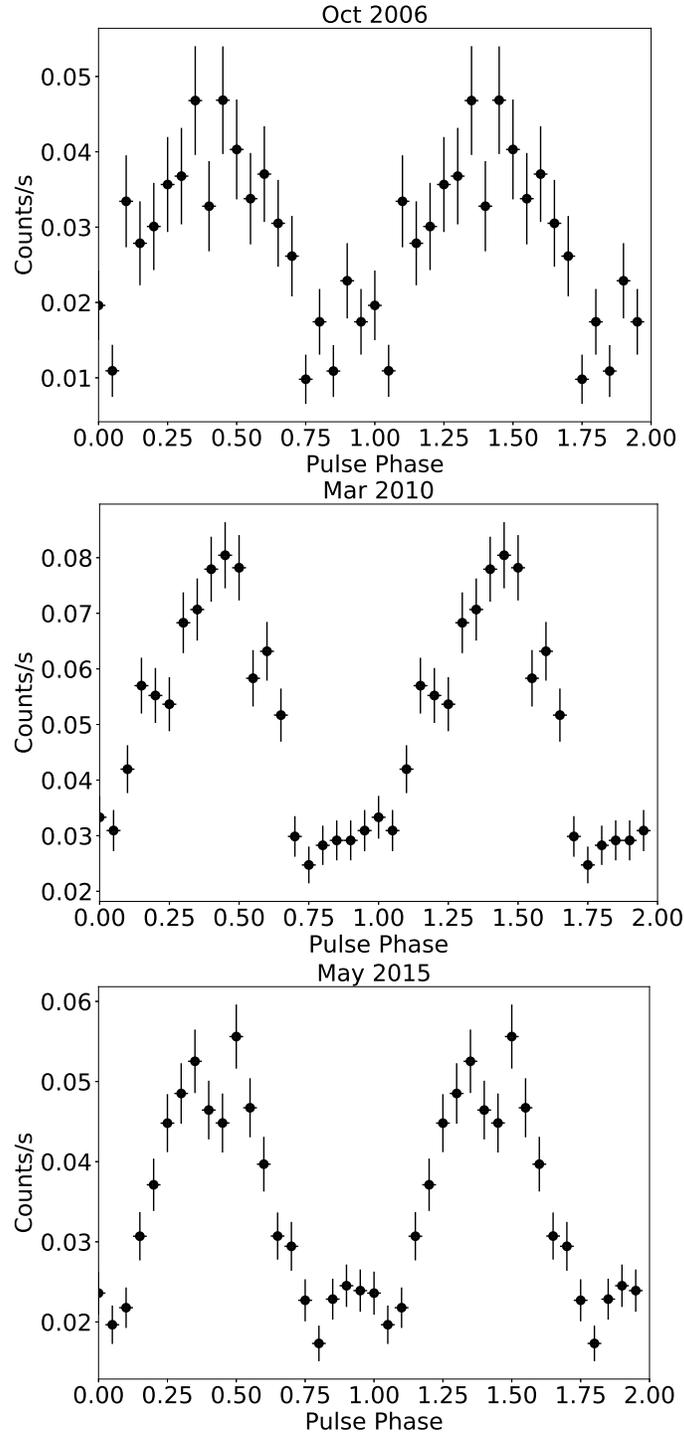}
\caption{Pulse profiles of 3XMM J181923.7$-$170616\ in October 2006, March 2010 and May 2015.}

\end{figure}
\label{f:pulse_profile}
}


The background-subtracted pulse profiles of 3XMM J181923.7$-$170616\ in 0.3--10 keV
are shown in Figure~\ref{f:pulse_profile}, in which one phase corresponds
to a period of 407.904 s.
The pulse profile maintains a nearly single-peak shape in 2006, 2010, 
and 2015, with small variation in the profile.

\subsection{Spectral Analysis} \label{sec:spec}

The \XMMN\ and \Swift\ spectra are extracted from 
a circular region centered at 3XMM J181923.7$-$170616\ with a radius of $30''$. 
The local background is selected from an annulus region centered at the point source with 
an inner radius of $55''$ and an outer radius of $120''$.
\begin{figure}[tbh!]
\centerline{
\includegraphics[angle=270, width=0.5\textwidth]{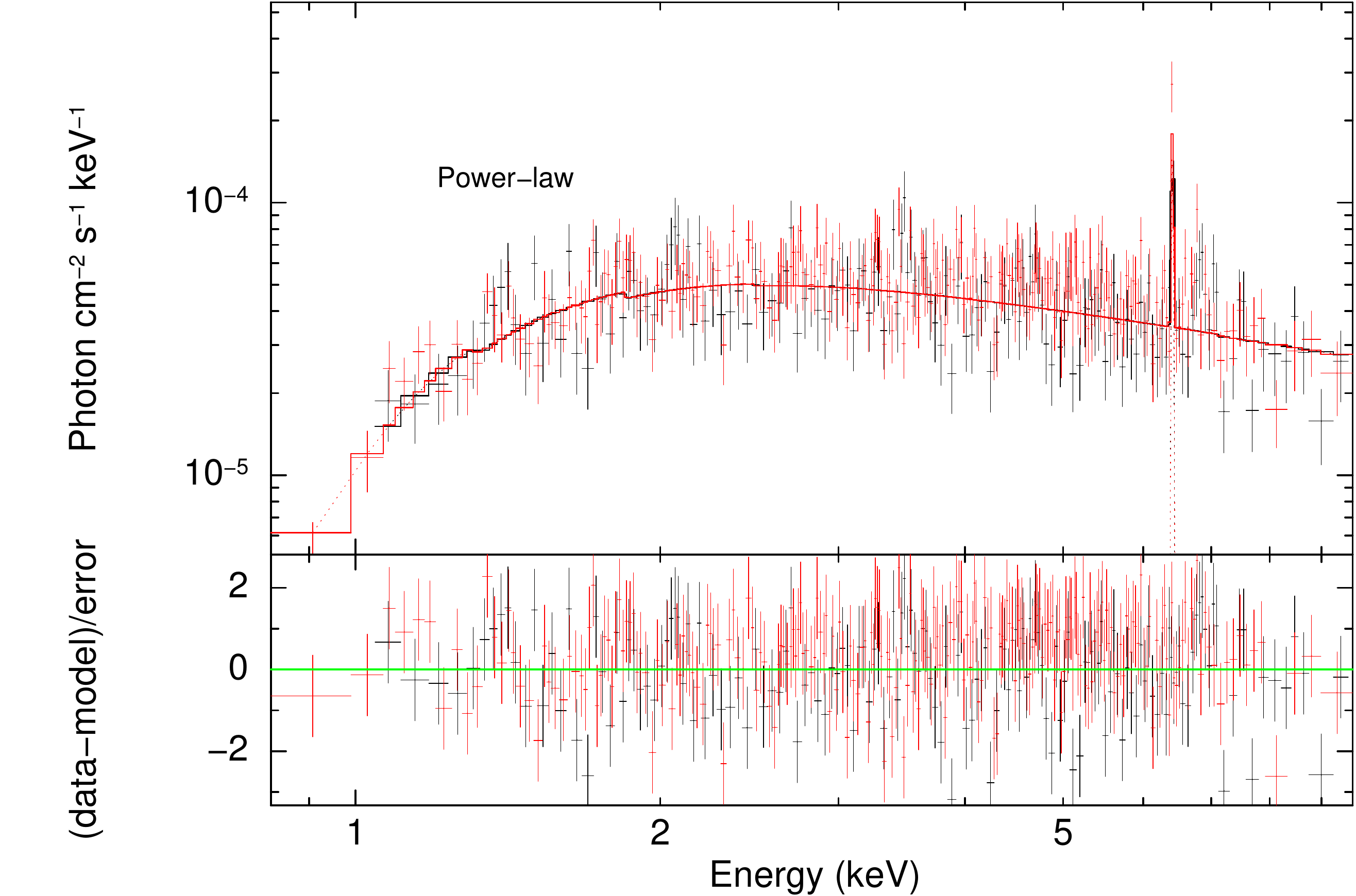}
\includegraphics[angle=270, width=0.5\textwidth]{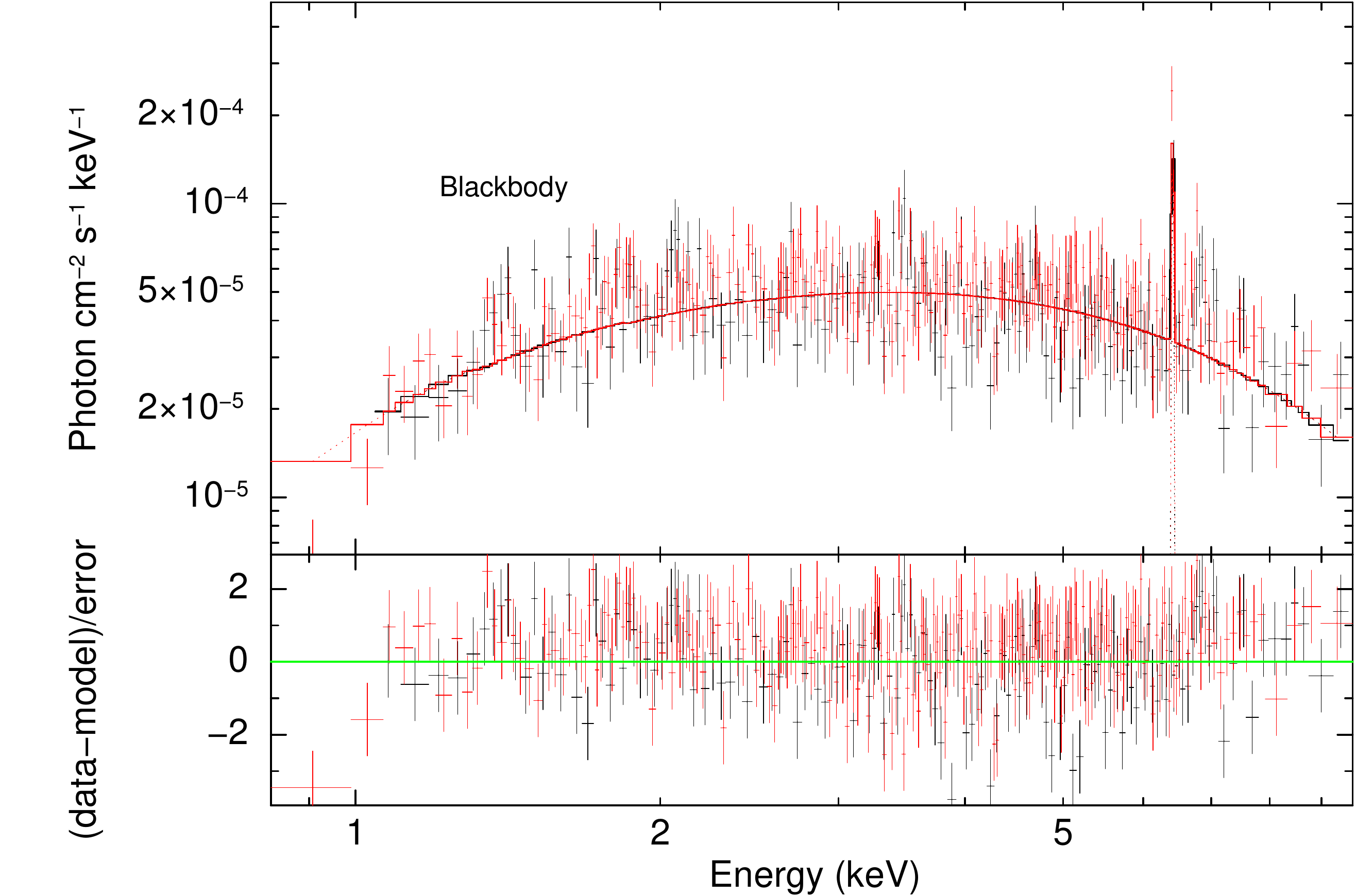}
}
\caption{{
Merged spectra of \XMMN\ MOS (black) and pn (red) fitted
  using an absorbed and \powerlaw\ model (left) or $blackbody$ model
  (right) with a Gaussian line at 6.4 keV. The fit results are shown 
  in Table~\ref{T:indiv_spec} .
 } 
}
\label{f:merged}
\end{figure}

\begin{figure}[tbh!]
\centering
\includegraphics[angle=0, width=0.9\textwidth]{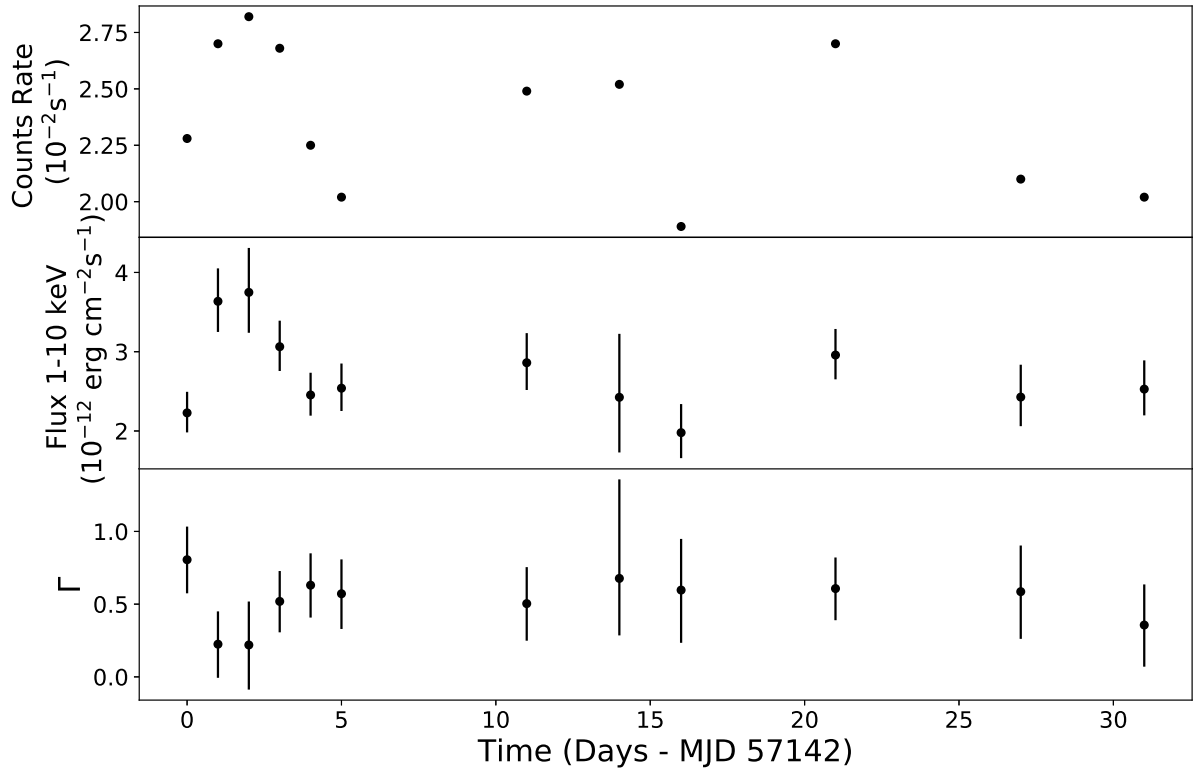}

\caption{ 
Variation of the counts rate, X-ray flux
and spectral index $\Gamma$ of
the 12 \Swift\ data taken in 2015.
}
\label{f:swiftxrt}
\end{figure}
In order to obtain an overall properties of 3XMM J181923.7$-$170616, 
{ first we produce two 
merged spectra of \XMMN\ 
(MOS1+2 and pn) by combining the long-time data of each instrument.
As shown in the left panel of Figure~3, the spectra can be well 
described by an absorbed \powerlaw\ model with a photon index 
of $\Gamma =0.58 \pm{0.07}$ plus a Gaussian line 
at $\sim 6.4$~keV ($\chi_\nu^2=1.11$).
The T\"{u}bingen-Boulder interstellar medium absorption model $tbabs$
is used for calculating the foreground absorption $\NH$ and the solar 
abundances are adopted from  Asplund et al.\ (2009).
The best-fit foreground absorption $\NH$ is $1.32^{+0.16}_{-0.15} \times 10^{22} \cm^{-2}$ 
(see spectral results in Table~\ref{T:indiv_spec}) obtained using merged \XMMN\ spectra.  
}
{ Although the fit can be slightly improved (${\chi_\nu^2}=1.10$) after 
adding a Gaussian line at 6.7 keV,
the line is too faint to be well constrained and is not clear 
in individual spectra.
Therefore, we only include the 6.4 keV iron $K\alpha$ emission for 
spectral fit.
We also model the spectra with an absorbed $blackbody$ model 
plus a 6.4 keV line, which gives an best-fit absorption of 
$\NH=0.33\pm{0.09} \times 10^{22} \cm^{-2}$ and a temperature of 
$kT=2.04^{+0.09}_{-0.08} \keV $. }
The $blackbody$ fit results in a ${\chi_\nu^2}=1.20$ slightly larger than 
the \powerlaw\ fit and displays large residuals $<1$ keV and $>$ 7 keV
(see the right panel of Figure~3).
Other models, such as $bremsstrahlung$ and $diskbb$, are tried but do not provide acceptable fit.


\begin{deluxetable}{lcccccc}
\tabletypesize{\footnotesize}
\tablecaption{Spectral fit results of the \XMMN\ observations}
\tablewidth{0pt}
\tablehead{
 & \multicolumn{5}{c}{Model 1:  \powerlaw}& \\
  \cline{2-6} 
Year   & $\chi_\nu^2/d.o.f.$  & $\NH$         &  $\Gamma$     &Fe norm$^a$  &Flux (1--10 keV)\\
  &                     & ($10^{22}~\cm^{-2}$)  &       & ($10^{-6}$cm$^{-2}$s$^{-1}$)  & ($10^{-12}$ $\erg \cm^{-2}$s$^{-1}$) 
 }
\startdata
Merged   &1.11/404 &$1.32_{-0.15}^{+0.16}$ & $0.58_{-0.07}^{+0.07}$ & $5.83_{-1.92}^{+1.92}$ &2.81$\pm{0.06}$
\\
2006  &0.99/116 &$1.63_{-0.46}^{+0.50}$ & $0.59_{-0.18}^{+0.19}$ & $-$ &2.83$\pm0.21$
\\
2010          &1.13/423 &$1.34_{-0.21}^{+0.24}$ & $0.59_{-0.21}^{+0.24}$ & $6.57_{-2.63}^{+2.64}$ &2.69$\pm0.13$
\\
2013  &0.90/91 &$1.11_{-0.42}^{+0.46}$ & $ 0.44_{-0.18}^{+0.19}$ & $-$ &3.44$ _{-0.27}^{+0.29}$
\\    
\cline{1-6} &\multicolumn{5}{c}{Model 2:  \it{blackbody}}&\\
\cline{2-6} 
Year         & $\chi_\nu^2/d.o.f.$ & $\NH$         &  $kT$     &Fe norm$^a$  &Flux (1--10 keV)\\     
&                   & ($10^{22}~\cm^{-2}$)  &  (keV)        & ($10^{-6}$cm$^{-2}$s$^{-1}$)  & ($10^{-12}$ $\erg \cm^{-2}$s$^{-1}$) \\
\hline
Merged       &1.20/402 &$0.33\pm{0.09}$ & $2.04^{+0.09}_{-0.08}$ & $6.11_{-1.9}^{+1.9}$ &2.48$\pm0.09$
\\
2006        &0.97/116 &$0.27_{-0.27}^{+0.38}$ & $2.06_{-0.20}^{+0.24}$ & $7.73_{-4.61}^{+4.63}$ &2.44$_{-0.20}^{+0.19}$
\\
2010        &1.17/423 &$0.33$(fixed) & $2.10_{-0.08}^{+0.08}$ & $1.92_{-1.20}^{+1.20}$ &2.56$_{-0.10}^{+0.10}$ 
\\
2013  &0.92/91 &$0.33$(fixed) & $2.13_{-0.17}^{+0.20}$ & $-$  &2.98$_{-0.25}^{+0.26}$
\\  
\enddata
\tablecomments{The errors are estimated at the 90\% confidence level.}
\tablenotetext{a}{\phantom{0} The flux of Fe line at 6.4 keV in $10^{-6}$ photons $\cm^{-2} \ps$.
The value is fixed to 0 and denoted with ``-'' when it can not be constrained.} 

\label{T:indiv_spec}
\end{deluxetable}


We subsequently model the spectra in each year to investigate the spectral evolution.
The \XMMN\ spectra in each year are jointly fitted (MOS1, MOS2 and pn).
We fit the spectra with the absorbed \powerlaw/$blackbody$ models plus a 
6.4 keV Gaussian line and show the best-fit results in Table~\ref{T:indiv_spec}.
The \powerlaw\ model generally provides a better fit than the $blackbody$ model, especially for 
describing the low-energy spectra.
The variation of $\NH$ can not be well constrained with the currently available data. 
{ The X-ray flux in 1--10 keV was $2.83\pm{0.21}\E{-12} \erg\cm^{-2}\ps$ 
in 2006, and was not significantly changed
in 2010.
The observation in 2013 was subject to flaring and we cannot further 
validate the nature of the flux change.
A short-term spectral variation was revealed from the 12 \Swift\ observation in 2015,
with the 1--10 keV flux increased by $1.52^{+0.5}_{-0.6}\E{-12} \erg\cm^{-2}\ps$ within 2 days (MJD 57142--57143; see Figure~\ref{f:swiftxrt}). 
The flux varied between $2.20\pm{0.14}\E{-12} \erg\cm^{-2}\ps$ and $3.75^{+0.56}_{-0.51}\E{-12} \erg\cm^{-2}\ps$,
with a hardening as the flux increased.
}



\begin{figure}[tbh!]
\includegraphics[angle=270, width=0.5\textwidth]{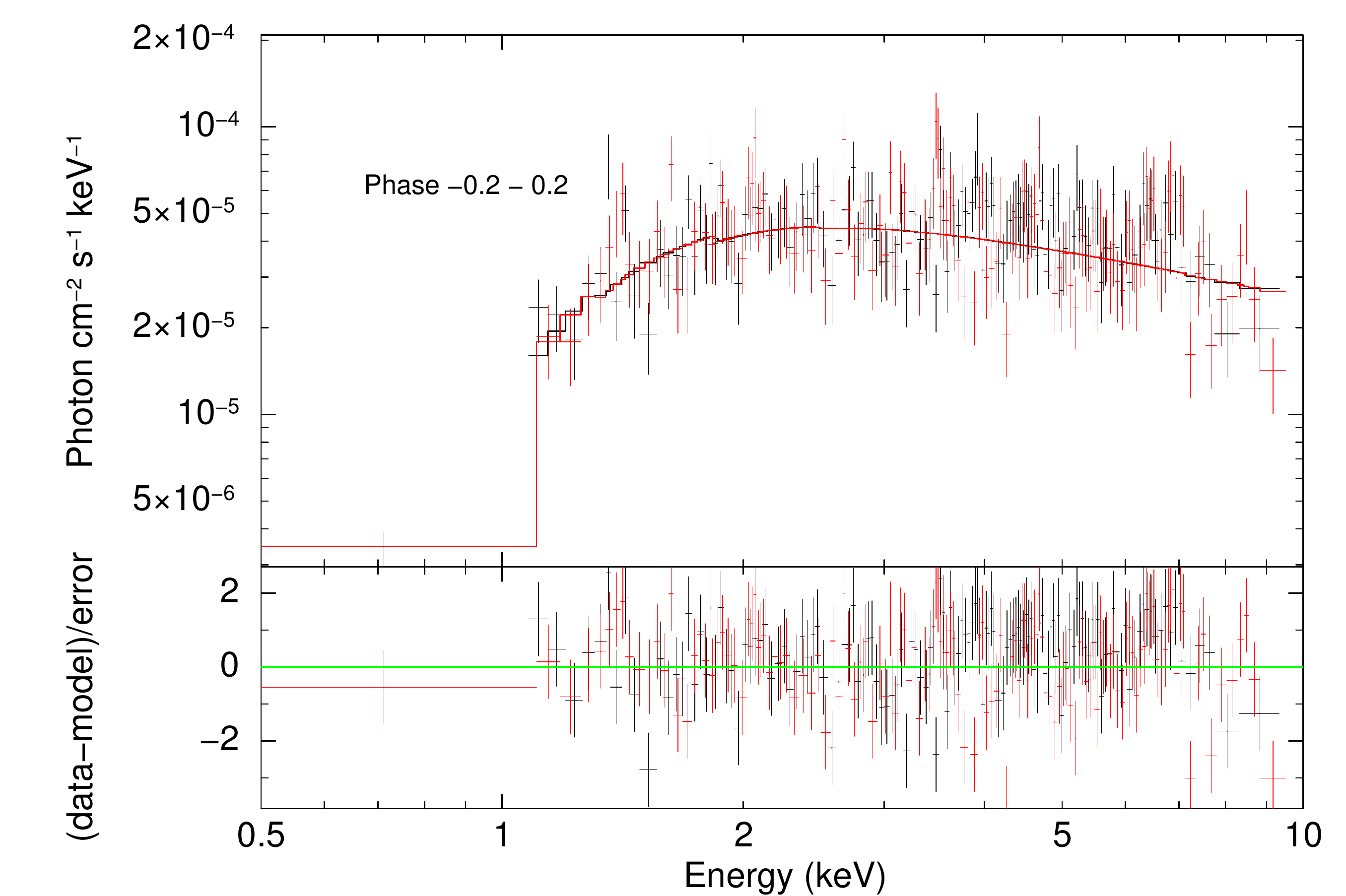}
\includegraphics[angle=270, width=0.5\textwidth]{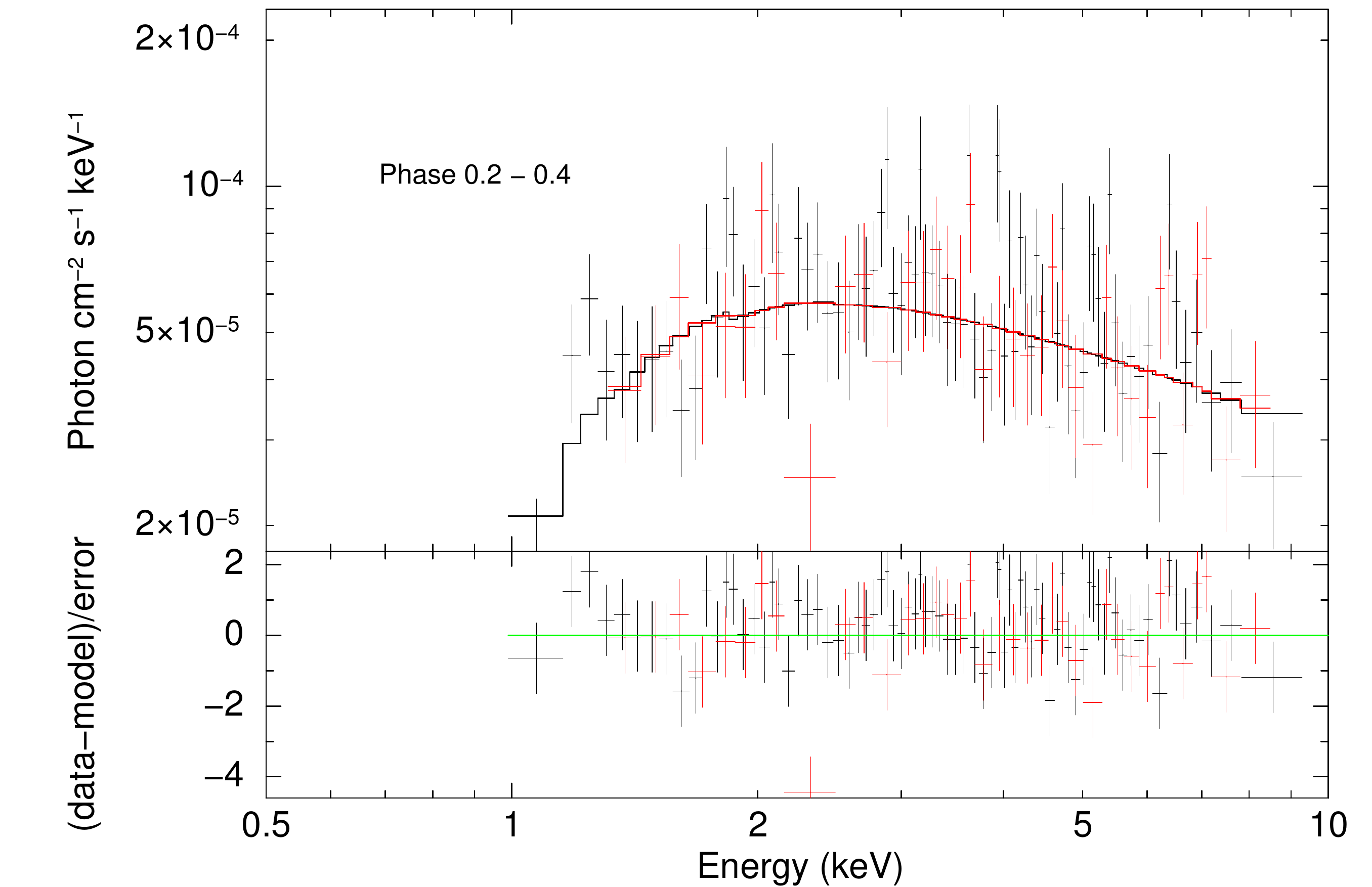}
\includegraphics[angle=270, width=0.5\textwidth]{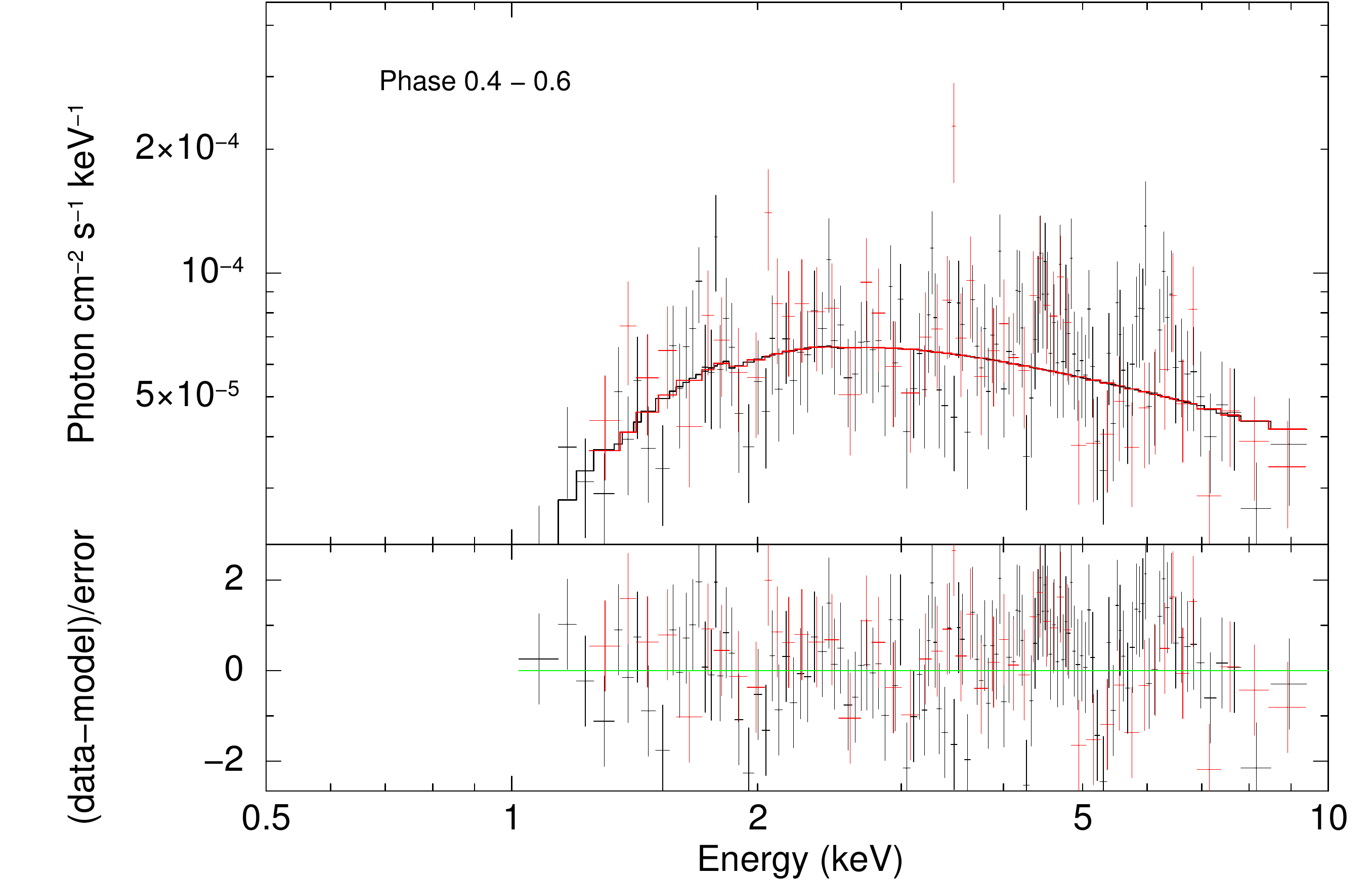}
\includegraphics[angle=270, width=0.5\textwidth]{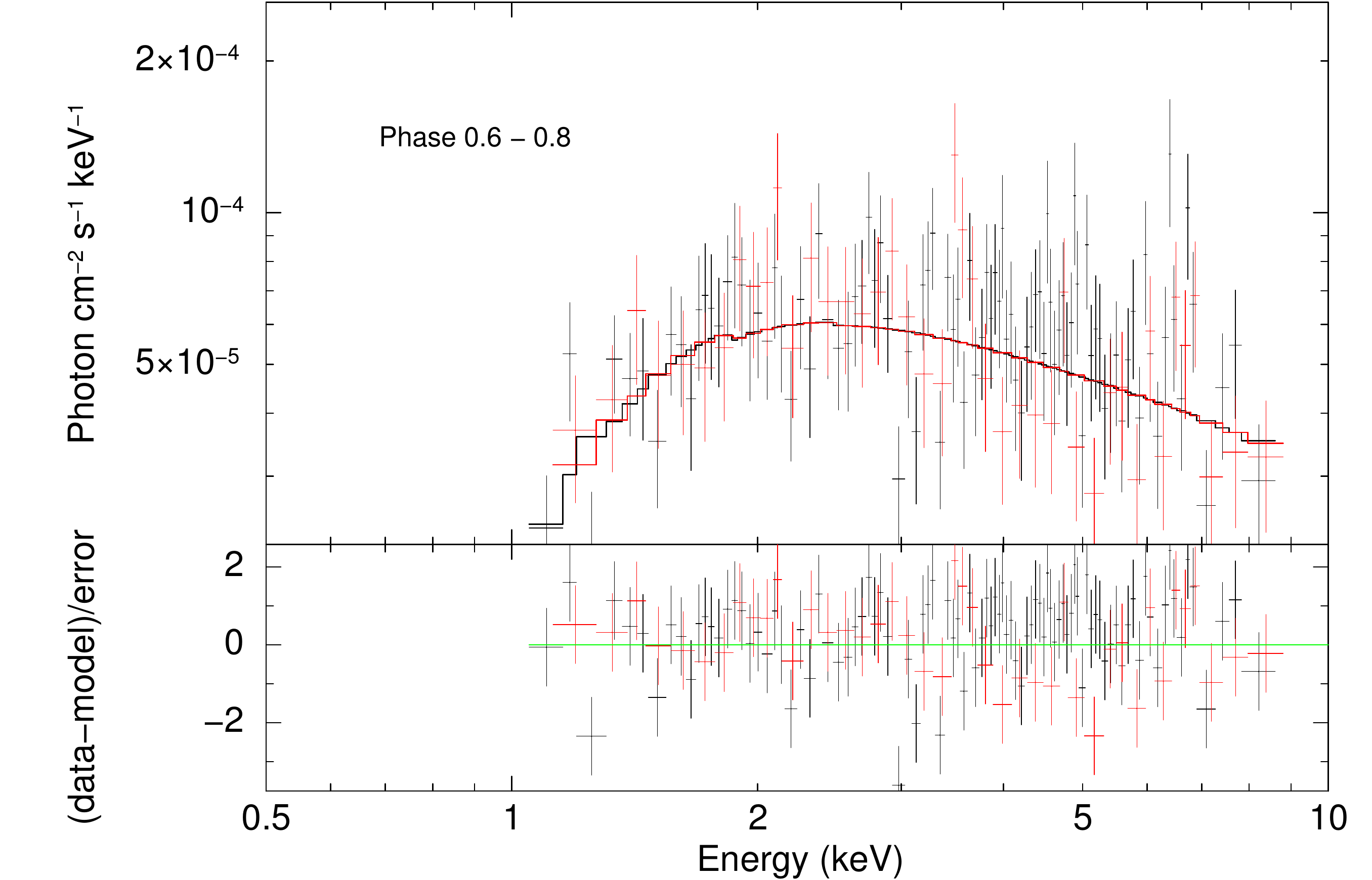}
\vspace{0.2in}
\caption{Phase-resolved spectra of \XMMN\ MOS (black) and pn (red) data 
fitted using an absorbed \powerlaw\ model. The fit results are shown 
in Table~3.
}
\label{f:phase}
\end{figure}

We also perform a phase-resolved spectroscopic analysis of the \XMMN\ spectra.
We divide the source time series of the \XMMN\ data into four individual phases: $-0.2$--0.2 (valley), 
0.2--0.4, 0.4--0.6 (peak), 0.6--0.8 phases (see Figure~\ref{f:phase}). 
We apply the absorbed \powerlaw/$blackbody$ models to jointly fit the 
spectra in the four phases (MOS 1 and MO2 are merged).
The fit results are shown in Table~3.

The absorbed \powerlaw\ model can best describe ($\chi_\nu^2 \lesssim 1.2$) the spectra of all phases
(see Figure~\ref{f:phase}), with an average 
column density of $\NH \sim1.2$--$1.3\times 10^{22} \cm^{-2}$ and a photon index  $\Gamma \sim0.6$.
Since the absorption value of phase 0.6--0.8 can not be constrained under the $blackbody$ model, we fix this value to the $\NH$ result of the merged spectra.
The table shows that among the phases there is little change to the parameters other than the flux.
We have also examined the Fe K$\alpha$ emission line flux in each phase, and 
have not found a significant change of the Fe line flux.
Hereafter, we adopt the \powerlaw\ results as they best describe the spectra.

\begin{deluxetable}{lccccc}
\tabletypesize{\footnotesize}
\tablecaption{Results of the phase resolved spectroscopy
with the \XMMN\ data}
\tablewidth{0pt}
\tablehead{
& \multicolumn{4}{c}{Model 1: {\it power-law} }\\
\cline{2-5}
Phase& $\chi_\nu^2/d.o.f.$ & $\NH$ &  $\Gamma$& Flux(1-10$\keV$)\\
&    & ($10^{22}~\cm^{-2}$) &  &($10^{-12}$\erg\ cm$^{-2}$s$^{-1}$)}
\startdata
-0.2 -- 0.2 & 1.22/269 & $1.25^{+0.24}_{-0.23}$ & $0.55^{+0.10}_{-0.10}$ & 2.56$\pm0.11$
\\
0.2 -- 0.4 
& 1.14/115 & $1.18^{+0.37}_{-0.32}$ & $0.59^{+0.16}_{-0.16}$ & 3.17$_{-0.23}^{+0.22}$
\\
0.4 -- 0.6 
& 1.15/173 & $1.29^{+0.30}_{-0.27}$ & $0.53^{+0.13}_{-0.13}$ &3.88$_{-0.22}^{+0.21}$
\\
0.6 -- 0.8
& 1.17/137 & $1.21^{+0.31}_{-0.30}$ & $0.62^{+0.16}_{-0.15}$&3.27$_{-0.21}^{+0.23}$
\\
\cline{1-5} &\multicolumn{4}{c}{Model 2:  \it{blackbody}}&\\
\cline{2-5}
Phase& $\chi_\nu^2/d.o.f.$ & $\NH$ & $kT$  & Flux(1-10$\keV$)\\
& & ($10^{22}~\cm^{-2}$) & (keV) &($10^{-12}$\erg\ cm$^{-2}$s$^{-1}$)\\
\hline
-0.2 -- 0.2
& 1.30/269 & $0.33$(fixed) & $2.15_{-0.09}^{+0.10}$ & 2.30$^{+0.10}_{-0.08}$
\\
0.2 -- 0.4 
& 1.20/115 & $0.14^{+0.22}_{-0.14}$ & $2.0\pm{0.2}$ & 2.74$_{-0.23}^{+0.21}$
\\
0.4 -- 0.6 
&  1.18/173 & $0.17^{+0.18}_{-0.17}$ & $2.2\pm{0.2}$ &3.47$_{-0.22}^{+0.21}$
\\
0.6 -- 0.8
& 1.30/137 & $0.33$ (fixed) & $1.9\pm 0.1$&2.78$_{-0.19}^{+0.20}$
\\
\enddata
\tablecomments{The errors are estimated at the 90\% confidence level.}
\label{T:resolve}
\end{deluxetable}

\section{Discussion} \label{sec:disc}

We have shown that 3XMM J181923.7$-$170616\ is a pulsating X-ray source with a period of 407.904(7)~s and a spin down
rate of $5.9\pm 5.4\E{-9}~\s \ps$.
The spectra is best characterized with a hard power-law with Fe lines at 6.4~keV and 6.7~keV.
{ 
The flux in 1--10 keV has varied between 2--$4\E{-12} \erg\cm^{-2}\ps$, and 
spectral index $\Gamma$ was in the range of 0.2--0.8.}
It appears to be a persistent source without luminous burst detected during the observation epochs.
The absorption column density $\NH=1.32^{+0.16}_{-0.15} \E{22} \cm^{-2}$ { determined by XMM observations} is similar to the value of the
Galactic hydrogen column density $\sim 1.2$--$1.3\E{22}~\cm^{-2}$ in this direction (Willingale \etal\ 2013).
It implies that the X-ray source is distant and possibly located near the 
boundary of the Galactic plane.
We conservatively set the lower limit of the distance of the X-ray source to 
the rotation tangent position at $\sim 8$~kpc. 
Hereafter the distance is parameterized as $10\dg$~kpc, given that 
the interstellar gas distribution along the line of sight is uncertain. 

\subsection{Search for optical and near-infrared counterpart} \label{S:counterpart}

We searched in the archival data and catalogues for the optical 
and near-IR counterparts of  3XMM J181923.7$-$170616. 
{ The 1$\sigma$ positional uncertainty for 90\% of 
the point sources in the 3XMM catalog is $2\farcs{4}$.}

\begin{figure}[tbh!]
\centering

\includegraphics[angle=0, width=0.5\textwidth]{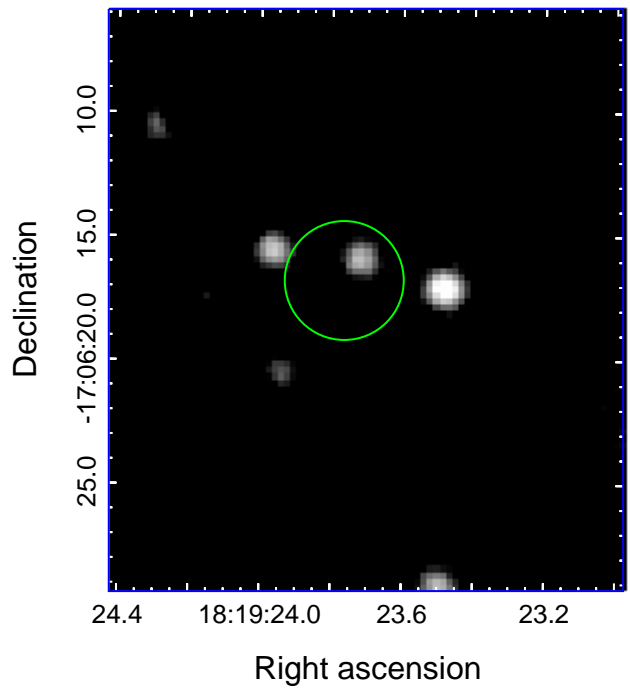}

\caption{I-band image obtained from the Pan-STARRS DR1 Archive. 
The green circle indicates the location of 3XMM J181923.7$-$170616\ and its 
$1\sigma$ uncertainty radius of $2\farcs{4}$.
}
\label{f:panstarrs}
\end{figure}

\begin{figure}[tbh!]
\centering
\includegraphics[angle=0, width=0.5\textwidth]{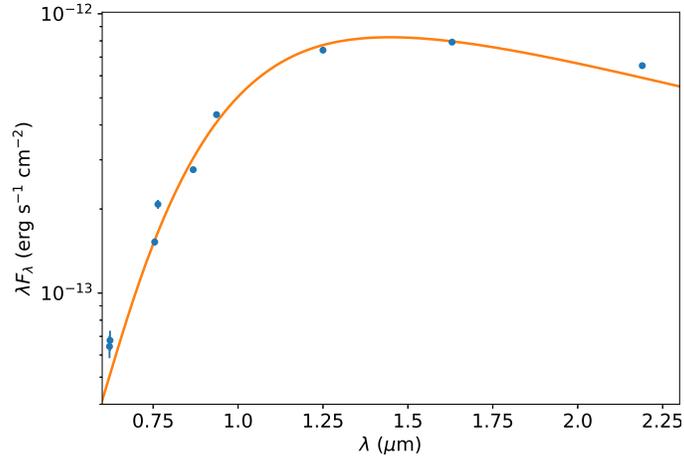}
\caption{
Spectral energy distribution of source \#1 (dots with error bars) fitted with
an absorbed blackbody model 
($A_{\rm V}=4.7$, $T_{\rm eff}=4226 \K$ and $L=4.7\E{34} \dg^2\erg\ps$).
}
\label{f:sed}
\end{figure}

Using the VizieR catalogue access tool (Ochsenbein \etal\ 2000), we found a faint optical 
source (source \#1; $\RA{18}{19}{23}{73}$, $\decldot{-17}{06}{16}{10}$, J2000) 
with the apparent magnitudes $M_R=20.90\pm0.09$ and $M_I=19.22\pm0.04$ in the VPHAS+ DR2 point 
source catalog (Drew \etal\ 2014, 2016), which is $0\farcs{5}$ away from 3XMM J181923.7$-$170616.
It also shows an H$\alpha$ line with a magnitude of $20.5\pm 0.2$. 
The optical source spatially matches a near-IR source ($M_J=16.758\pm0.018$, $M_H=15.939\pm0.019$, 
and $M_K=15.320\pm0.026$,) recorded in the UKIDSS Galactic Plane Survey (Lucas \etal\ 2012).
We also notice that it matches the source PSO J181923.731-170616.083 in the Panoramic Survey Telescope and Rapid Response System (Pan-STARRS) survey 
(Chambers \etal\ 2016).
Source \#1 was detected in the r,i,z,y bands among the five filter bands used in the Pan-STARRS survey.
Figure~\ref{f:panstarrs} shows the I-band image of the optical source.
In Figure~\ref{f:sed}, we show the spectral energy distribution (SED) 
of source \#1. It includes photometry in the bands mentioned above.
We fit the SED with an absorbed blackbody model, while the absorbed power-law model
can not explain the convex spectra.
The V-band absorption $A_V$ of $4.7$ is converted from the foreground absorption 
$\NH=1.35 \times 10^{22}$ cm$^{-2}$ using a conversion factor 
$\NH=2.87 \times10^{21} A_V$ cm$^{-2}$ (Foight \etal\ 2016).
The ratio of total to selective extinction at V band $R_V=3.1$ is adopted for the
calculation of the absorption at each wavelength.
The fit shows that source \#1 has an effective temperature of $T_{\rm eff}
=4226\pm151 \K$ and luminosity of $4.7\pm0.2\E{34} \dg^2 \erg\ps$.
The temperature and luminosity of source \#1 is similar to a giant star
with a spectral type of K3 III.

There is another near-IR UKIDSS source located $\sim 2"$ away (source \#2; 
$\RAdot{18}{19}{23}{76}$, $\decldot{-17}{06}{17}{67}$,
J2000) with $M_J=17.497\pm 0.035$, $M_H=15.547\pm0.014$, and 
$M_K=14.622\pm0.014$, which might be the same source with $M_{3.6\um}=13.5\pm0.1$, 
$M_{4.5\um}=13.76\pm 0.30$, and $M_{5.8\um}=12.34\pm 0.29$ in the GLIMPSE source catalog.
The blackbody fit shows that source \#2 has a temperature of $\sim 1880~\K$ and 
luminosity of $3.5\E{34} \dg^2 \erg\ps$.

The spectra of source \#1 and \#2 are not compatible with 
OB supergiants ($10^4 L_\odot$) or hot main-sequence stars ($T_{eff}>10^4 \K$) 
Nevertheless, we can not exclude the possibility that the donor 
star of 3XMM J181923.7$-$170616\ is a Be and the emission of the stellar component was not 
detected with current surveys.
Some Be stars exhibit IR excess with a blackbody temperature of thousands 
of kelvin, which is attributed to the circumstellar material (Zhang 
\etal\ 2005).
Assuming a detection limit of 21 magnitude in V band, current surveys may fail to detect 
a star with an absolute magnitude $>-1.1$ at an assumed distance of 30 kpc (near Galactic edge;
$A_V=4.7$), which corresponds to a main-sequence star no earlier than B3-type.
It is also unclear whether the two IR sources are relevant or only 
foreground/background sources projected near 3XMM J181923.7$-$170616. 
Further targeted optical and IR observations are needed to achieve a firm conclusion.

{ The position accuracy of 3XMM J181923.7$-$170616\ can be improved by performing an
astrometric analysis with the 
Swift-XRT data products generator\footnote{http://www.swift.ac.uk/user\uline{ }objects/}.
The position coordinates are derived from the XRT data product detecting and localizing all sources in the image, and then matching them with the 2MASS catalogue sources (Evans \etal\ 2014).
This substantially increases the position accuracy 
to a 90\% confidence error radius of
$0\farcs{6}$ centered at $\RAdot{18}{19}{23}{73}$, $\decldot{-17}{06}{15}{9}$, which
is consistent with the position of the optical--IR source \#1.
}

\subsection{The nature of 3XMM J181923.7$-$170616}


3XMM J181923.7-170616 was considered to be a candidate of a slowly spinning X-ray pulsar 
(Farrell et al. 2015) and its nature remained unclear.
Based on the analysis described above, we here discuss three possible scenarios for the nature 
of the X-ray source: an isolated NS, a cataclysmic variable (CV) or an X-ray binary.


{ If 3XMM J181923.7$-$170616\ is an isolated NS, the long period and the period derivative
would suggest 
a rotational energy loss $\dot{E}_{\rm rot}=3.95\E{46} P^{-3} 
\dot{P}\leq 3.4\pm 3.1 \E{30} \erg\ps$, which
is insufficient to power the observed X-ray emission ($L_X=2.78\times 10^{34} \dg^2 \erg\ps$). }
In this case, { the long period and low $\dot{E}_{\rm rot}$ would suggest
that the source is not a classical rotation powered NS, but may be a magnetar.}
Magnetars are NSs with ultra-strong magnetic fields and their 
X-ray emission is powered by the decay of the magnetic fields 
(Thompson \& Duncan 1995, 1996; Thompson et al. 2002). 
The known magnetars have periods 2-12 s, except one candidate 
1E 161348$-$5055 in the supernova remnant RCW 103 
showing surprisingly large period of 6.67 hr (De Luca et al. 2006). 
Besides the anomalous long period, the X-ray emission of 3XMM J181923.7$-$170616\ appears to 
be much more stable than { magnetars undergoing
outbursts, when the magnetars are up to $10^3$
times brighter than the steady state and then experience a decay lasting a few weeks to months (Rea \& Esposito 2011).}
It also can not be a quiescent magnetar, since the quiescent X-ray emission 
of a mangnetar is soft according to 
the magneto-thermal evolution model (Vigan{\`o} \etal\ 2013) and observations
(a power-law photon index = 1.5-4 or blackbody temperature $kT_{BB}$ = 0.1-0.7 keV; Olausen \& Kaspi 2014).
Moreover, the iron line at 6.4 keV supports the existence of cold materials surrounding the source. 
Therefore, the isolated pulsar explanation is not favored according to the spectral behaviors.

3XMM J181923.7$-$170616\ is more luminous than CVs which have a typical luminosity $L_X\lesssim 10^{34} \erg\ps$ (Burenin \etal\ 2016). 
CVs show blackbody radiation during the outburst state (Mukai \etal\ 2003), while in 
quiescent state the emission becomes optically thin and presents multi-temperature 
Bremsstrahlung radiation (Mukai 2001, Richman \etal\ 1996, Bernardini \etal\ 2012).
The X-ray emission of 3XMM J181923.7$-$170616\ does not show significant variation and the spectra 
cannot be well fitted with an optically thin model, making the CV explanation unlikely.

The remaining possibility is an X-ray binary.
Since only one periodicity has been found with the \Swift\ and \XMMN\ data,
we need to discuss whether the 408 s is a spin period or an orbital period.
According to the Kepler's third law, the orbital period $P_{\rm orb}$ 
depends on the masses of the primary ($M_{\rm X}$) and the donor 
($M_*$), and the separation between them ($a$):
$P_{\rm orb}=2\pi a^{3/2}/[G(M_{\rm X}+M_{\rm *})]^{1/2}$.
Assuming $P_{\rm orb}= 408~\s$ and $M_{\rm X} = 1.4~\Msun$ for an NS, the separation 
$a$ are obtained as $0.16 R_\odot$ and $0.27 R_\odot$, respectively (for a donor mass of 
$1~\Msun$ and $10~\Msun$, where $R_\odot$ is the solar radius).
If the 408~s is an orbital period, the companion radius must be much 
smaller than solar radius (e.g., a degenerate companion), and the 
X-ray source could be a ultra-compact LMXB with the smallest known 
$P_{\rm orb}$ among its group.
However, the hard spectra ($\Gamma<1$ or $kT\sim 2$~keV) are atypical for 
those low-level accretion LMXBs ($L_X\sim 10^{34-36}~\erg\ps$ and 
$\Gamma\gtrsim 1.4$; e.g., Wijnands \etal\ 2015; 
in't Zand \etal\ 2005; Wijnands \etal\ 2006) or for those in the
quiescent (temperature of less than a few hundred eV; 
$L_X=10^{30-33}\erg\ps$; e.g., Heinke \etal\ 2003).
Therefore, we suggest the 408 s is the pulsar spin period.

The long spin period and hard X-ray spectra widely exist in HMXBs.
The hard power-law emission can be interpreted as Comptonization of the thermal
photons by the high-energy electrons.
3XMM J181923.7$-$170616\ is unlikely to be a SGXB due to the absence of OB supergiant 
at its location (see Section~\ref{S:counterpart}).
The X-ray properties of the source indeed well matches the characteristics of persistent 
Be X-ray binaries, including the long pulse period ($P_{\rm spin} 
\gtrsim 200~\s$), persistent 
and low-luminosity ($\le 10^{34-35} \erg\ps$), low X-ray variability and hard spectrum with faint 
Fe 6.4~keV line (Reig \& Roche 1999; Reig 2011).
In this case, the NS is orbiting in a low-density region of the Be star's wind.
This explanation is waiting to be tested, given that no massive star 
earlier than B3 has been detected at the position of 3XMM J181923.7$-$170616.

SyXBs are a new group of XRBs that display some X-ray 
properties similar to HMXBs but are systems that host a K/M-type 
giant as its donor. 
These objects display long pulse periods ($\gtrsim 100 \s$), low 
luminosity, and hard power-law spectra (0.5--2; Enoto \etal\ 2014). 
As mentioned in Section~\ref{S:counterpart}, an optical/IR source spatially 
consistent 3XMM J181923.7$-$170616\ shows characteristics of an K3 giant.
The SyXBs scenario is therefore most favorable nature of 3XMM J181923.7$-$170616\
due to the existence of a late type counterpart candidate.

Hence, 3XMM J181923.7$-$170616\ is most likely an SyXB with a K-type giant donor (source
\#1, see Figure~\ref{f:panstarrs}).
An alternative possibility is that it is a persistent BeXB with 
a companion star no earlier than B3.
Future X-ray and optical monitoring observations will shed light on the other 
period and the type of the companion star, so as to confirm the nature of 3XMM J181923.7$-$170616.

\section{Summary}

We have performed a detailed X-ray analysis of 3XMM J181923.7$-$170616\ using the \XMMN\ and 
\Swift\ observations spanning over 9 years.
The main conclusions are the following:

\begin{enumerate}
\item{We have accurately determined the spin period of $P=407.904(7)\s$ and {discovered
the spinning down of the source with $\dot{P}\leq 5.9\pm 5.4 \E{-9} \s \ps$
($1\sigma$).}
The pulse shape is similar to a sinusoid profile and does show significant change in 9 yrs.
}

\item{3XMM J181923.7$-$170616\  emits persistent X-ray emission which is best characterized {by an 
absorbed power-law emission with $\NH\sim 1.32\E{22}~\cm^{-2}$ and $\Gamma \sim0.6$
plus two Fe lines at 6.4 keV and 6.7 keV.
The source experienced a small flux variation  (2--$4\E{-12} \erg\cm^{-3} \ps$), with a spectral hardening as the flux increased.
}
No burst activities have been observed during the observation epochs.
We performed phase-resolved spectroscopy and do not find significant change
of $\NH$, $\Gamma$ and the flux of 6.4~keV line between different phases.
}

\item{The absorption column density of 3XMM J181923.7$-$170616\ is similar to the total Galactic $\NH$
along its direction, indicating that it is a distant source.
We searched for optical and IR counterparts from the archival surveys, and exclude
the existence of a Galactic OB supergiant. 
We discover an optical counterpart with a temperature and luminosity similar to a K3-type giant.}

\item{We discussed the nature of 3XMM J181923.7$-$170616\ by comparing its properties with those of isolated
NSs, CVs and X-ray binaries. 
{ It is unlikely to be an isolated magnetar, given the relatively small variability, hard spectra, and the existence of surrounding cold materials as indicated by the 6.4 keV line.} 
The luminosity and the spectra are also not consistent with the properties of a CV.
An X-ray binary is the probable explanation.
An SyXB is the favored nature of 3XMM J181923.7$-$170616\ and can essentially explain its low luminosity, 
slow pulsation, hard spectrum, and possible late type companion. 
An alternative explanation of the source is a persistent Be/X-ray 
binary with a companion star no earlier than B3-type.
}

\end{enumerate}

\begin{acknowledgements}
{ We acknowledge the anonymous referee for suggestions regarding the phase-connection 
analysis and the position accuracy.
We thank Peng Wei and Eugene Magnier for helpful discussion about the search of the 
optical/IR counterpart.
This work was supported by} NSFC grants 11503008 and J1210039,
and Top-notch Academic Programs Project of Jiangsu Higher Education Institutions.
P.Z. acknowledges the support from the NWO Veni Fellowship, grant no.\ 639.041.647, { and NSFC grant 11233001.}
W.Y. was supported in part by the NSFC grant no.\ 11333005 and by the National Program 
on Key Research and Development Project (grant no.\ 2016YFA0400804).
X.L. was supported by the National Program on Key Research and Development Project (grant no.\
2016YFA0400803), and the NSFC grants 11133001 and 11333004.
X.X. thanks the support by NSFC grant 11303015.

\end{acknowledgements}


\end{document}